\documentclass[twocolumn]{revtex4}
\usepackage{graphicx,epsfig}
\usepackage{amsmath}
\usepackage{amsfonts}
\usepackage{xcolor}
\usepackage{fancyhdr}
\usepackage{hyperref}
\usepackage{lipsum}
\usepackage{tikz}
\usetikzlibrary{arrows.meta,positioning}
\DeclareUnicodeCharacter{2212}{-}

\begin{document}




\title{Correlation-Based Diagnostics of Social Contagion Dynamics in Multiplex Networks}
\author{Joan Hernàndez Tey}
\affiliation{Facultat de F\'{\i}sica, Universitat de Barcelona, 08028 Barcelona, Spain.}

\author{Emanuele Cozzo}
\affiliation{Departament de F\'isica de la Mat\`eria Condensada, Facultat de F\'{\i}sica, Universitat de Barcelona, 08028 Barcelona, Spain.\\ Universitat de Barcelona Institute of Complex Systems (UBICS), Universitat de Barcelona, 08028 Barcelona, Spain. \\ Communication Networks \& Social Change research group, Universitat Oberta de Catalunya, Barcelona, Spain.}

\begin{abstract}
Multiplex contagion dynamics display localization phenomena in which spreading activity concentrates on a subset of layers, as well as delocalized regimes where layers behave collectively. We investigate how these regimes are encoded in temporal correlations of node activity. By deriving a closed-form mean-field expression for node autocorrelations in a contact-based social contagion multiplex model and validating it through simulations, we show that lag-one autocorrelations act as sensitive indicators of both activation and localization transitions. Our results establish temporal correlations as lightweight, structure-agnostic probes of multiplex spreading dynamics, particularly valuable in partially observable systems.
\end{abstract}

\maketitle
\section{Introduction}

In contemporary digital ecosystems, social activity, information diffusion, and collective behavior unfold simultaneously across multiple online platforms. Individuals typically maintain accounts on several services, interacting through distinct but interdependent networks. As a result, many dynamical processes of interest --- ranging from opinion formation and attention cascades to coordination and collective mobilization --- are intrinsically \emph{multiplex} in nature. Understanding such processes therefore requires models and analytical tools that explicitly account for cross-platform coupling rather than treating each platform in isolation.

A central challenge in this context is \emph{partial observability}. In real-world settings, access to data is often asymmetric: some platforms provide extensive public data, while others impose strong restrictions or are entirely opaque. Consequently, researchers and practitioners are frequently faced with the problem of inferring the \emph{global state} of a coupled, multi-platform system using observations from only a subset of its layers. This raises a fundamental question: \emph{to what extent can the macroscopic behavior of a multiplex dynamical process be diagnosed from limited, layer-specific observations?} Addressing this question is crucial both for empirical analysis and for the design of timely interventions in networked systems where direct monitoring of all components is infeasible.

In this work, we approach this inference problem through the lens of \emph{contact-based contagion dynamics on multiplex networks}. While epidemic-like models have long been used to study spreading processes on single networks, their extension to multiplex settings reveals qualitatively new regimes of behavior. In particular, depending on the strength of interlayer coupling and the structural heterogeneity between layers, contagion may remain \emph{localized}, with one layer effectively driving the dynamics, or become \emph{delocalized}, with multiple layers activating simultaneously. Previous studies have shown that this localization--delocalization crossover can be predicted from structural properties of the multiplex network alone. Here, we go one step further by asking whether \emph{dynamical correlations}, accessible from time series data on individual layers, can serve as reliable indicators of the global regime of the system, even when only a single layer is observed.

To this end, we focus on \emph{node-level temporal autocorrelations} as a minimal yet informative observable. Autocorrelations are directly accessible from time-resolved activity data, require no knowledge of the full network structure, and naturally encode the balance between endogenous spreading and externally driven fluctuations. We introduce a mean-field analytical approximation for the steady-state autocorrelation in a contact-based multiplex contagion model and analyze how its behavior reflects the main regions of the multiplex phase diagram. By combining analytical results with dynamical simulations, we show that autocorrelations provide clear and robust signatures distinguishing inactive, active--localized, and active--delocalized regimes, and we assess the conditions under which the state of an unobserved layer can be inferred from partial observations. The paper is organized as follows: in Sec.~\ref{sec:model} we introduce the model and summarize its phase diagram; in Sec.~\ref{sec:autocorrelation_approx} we derive the analytical approximation for node autocorrelations; in Sec.~\ref{sec:results} we validate and interpret these predictions through numerical simulations; and we conclude with a discussion of implications and limitations for inference in partially observable multiplex systems.

\section{The model}
\label{sec:model}
In this work we use the contact-based social contagion model on multiplex networks introduced in Ref.~\cite{PhysRevE} and further analyzed in Ref.~\cite{tey2024eigenvectorlocalizationuniversalregime} as a minimal framework for modeling propagation of information across multiple, interdependent social platforms. The model incorporates both intra-layer spreading within each platform and interlayer transmission between replicas of the same user across platforms. Since our focus here is on dynamical autocorrelations, we only summarize the main ingredients of the model and its qualitative regimes, referring the reader to Refs.~\cite{PhysRevE,tey2024eigenvectorlocalizationuniversalregime} for full details and derivations.

We consider a multiplex network composed of $M$ layers, each containing the same set of $N$ nodes. Following the formalism introduced in \cite{cozzo2018multiplex}, latin indices $u,v \in \{1,\dots,N\}$ label nodes within a layer, while Greek indices $\alpha,\kappa \in \{1,\dots,M\}$ denote layers. Global indices $i,j \in \{1,\dots,MN\}$ refer to nodes in the full multiplex and are defined through the bijection $i = u\alpha$.

The model interpolates between the contact process (CP), where an infected node makes at most one effective contact per time step, and the reactive process (RP), where it attempts to infect all its neighbors in the layer at each time step. For each layer $\alpha$, the contact probability matrix $R^\alpha$ is defined as in the single layer case introduced in \cite{G_mez_2010}:
\begin{equation}
(R^\alpha)_{uv} = 1 - \left(1 - \frac{(A^\alpha)_{uv}}{k_u^\alpha} \right)^{\gamma_u^\alpha},
\label{eq:Ralpha}
\end{equation}
where $A^\alpha$ is the adjacency matrix of layer $\alpha$, $k_u^\alpha$ is the degree of node $u$, and $\gamma_u^\alpha$ controls the interpolation between CP ($\gamma=1$) and RP ($\gamma \to \infty$) dynamics.

The full multiplex dynamics is encoded in the supra-contact probability matrix
\begin{equation}
\bar{R} = \bigoplus_{\alpha=1}^M R^\alpha + \epsilon\, C,
\label{eq:Rbar}
\end{equation}
where $C$ is the interlayer coupling matrix that connects each node to its replicas across layers, and $\epsilon = \eta/\beta$ is the ratio between the interlayer infection probability $\eta$ and the intralayer infection probability $\beta$.

Let $p_i(t)$ denote the probability that node $i$ is infected at time $t$. Under a mean-field approximation that neglects joint infection probabilities, the discrete-time evolution equation reads~\cite{PhysRevE, G_mez_2010}
\begin{equation}
p_i(t+1) = [1 - p_i(t)] [1 - q_i(t)] + (1 - \mu) p_i(t) + \mu [1 - q_i(t)] p_i(t),
\label{eq:mstereq}
\end{equation}
where $\mu$ is the recovery probability and
\begin{equation}
q_i(t) = \prod_j \left(1 - \beta \bar{R}_{ij} p_j(t) \right)
\label{eq:qi}
\end{equation}
is the probability that node $i$ does not receive infection from any neighbor at time $t$.

At stationarity, the absorbing solution $p_i = 0$ becomes unstable when
\begin{equation}
\frac{\beta}{\mu} > \frac{1}{\bar{\Lambda}_{\max}},
\label{eq:threshold}
\end{equation}
where $\bar{\Lambda}_{\max}$ is the largest eigenvalue of the supra-contact matrix $\bar{R}$. The critical point separating the inactive and active phases is therefore given by
\begin{equation}
\left( \frac{\beta}{\mu} \right)_c = \frac{1}{\bar{\Lambda}_{\max}}.
\label{eq:critical_point}
\end{equation}

\begin{figure}[t]
\centering
\begin{tikzpicture}[x=6.2cm,y=4.6cm,>=Latex, font=\small]

  \draw[->, line width=0.9pt] (0,0) -- (1.05,0)
    node[below right] {$\beta/\mu$};
  \draw[->, line width=0.9pt] (0,0) -- (0,1.05)
    node[above left] {$\eta/\beta$};

  \def\xb{0.38}   
  \def\xn{0.72}   
  \def\yc{0.58}   
\fill[blue!12] (0,0) rectangle (\xb,1);                 

\fill[orange!18] (\xb,0) rectangle (\xn,\yc);           

\fill[green!18] (\xn,0) rectangle (1,\yc);              

\fill[red!15] (\xb,\yc) rectangle (1,1);                
  \draw[densely dashed, line width=0.9pt] (\xb,0) -- (\xb,1)
    node[pos=0.02, below, xshift=-2pt] {\scriptsize $1/{\bar{\Lambda}_{max}}$};

  \draw[densely dashed, line width=0.9pt] (\xn,0) -- (\xn,\yc)
    node[pos=0.02, below, xshift=-2pt] {\scriptsize $1/\Lambda^{(2)}$};

  \draw[densely dashed, line width=0.9pt] (\xb,\yc) -- (1,\yc)
    node[pos=0.5, above] {\scriptsize $(\eta/\beta)_t \sim \Lambda_1-\Lambda_2$};

  \node[align=center] at (0.18,0.5) {\bf I\\[-1mm]\scriptsize inactive};

  \node[align=center] at (0.55,0.25) {\bf AL\\[-1mm]\scriptsize active\\\scriptsize localized};

  \node[align=center] at (0.86,0.25) {\bf AD$_1$\\[-1mm]\scriptsize active\\ \scriptsize delocalized};

  \node[align=center] at (0.72,0.82) {\bf AD$_2$\\[-1mm]\scriptsize active delocalized};

  \draw[->, line width=0.6pt] (0.50,0.50) -- (0.50,0.70);
  \node[align=center] at (0.52,0.52) {\scriptsize increase\\ \scriptsize coupling};

  \draw[->, line width=0.6pt] (0.52,0.14) -- (0.80,0.14);
  \node[align=center] at (0.63,0.10) {\scriptsize increase transmissibility};

\end{tikzpicture}
\caption{Schematic phase diagram in the plane of transmissibility $\beta/\mu$ and interlayer coupling $\eta/\beta$ for a two-layer multiplex with $\Lambda_1>\Lambda_2$. The absorbing phase (I) lies to the left of the criticalpoint $\beta/\mu=1/\bar{\Lambda}_{max}$. In the active phase, strong coupling above $(\eta/\beta)_t\sim \Lambda_1-\Lambda_2$ yields delocalized activity (AD$_2$). For weaker coupling, increasing $\beta/\mu$ above the non-dominant activation threshold $(\beta/\mu)_c^{(2)}$ drives the system from an active--localized regime (AL) to a delocalized endemic regime (AD$_1$). The diagram is schematic and not to scale.}
\label{fig:phase_diagram_schematic}
\end{figure}
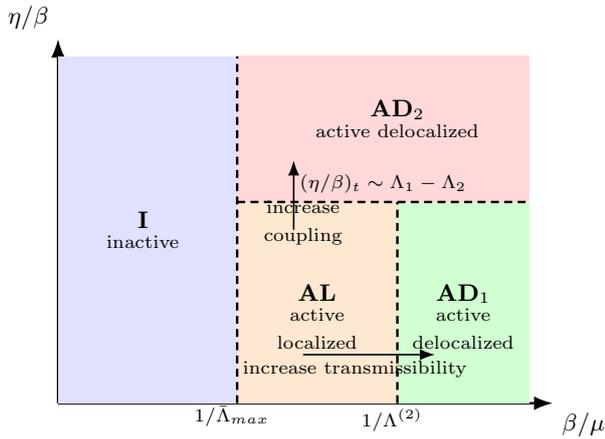

The interplay between intra- and interlayer couplings gives rise to two qualitatively distinct spreading regimes~\cite{tey2024eigenvectorlocalizationuniversalregime,Ferraz_de_Arruda_2020,de2017disease}:

\begin{itemize}
\item \textbf{Localized regime (AL).}
As illustrated in Fig.~\ref{fig:phase_diagram_schematic}, the localized regime corresponds to the region labeled AL in the phase diagram. In this regime the multiplex is endemic because the dominant layer (the one with the largest leading eigenvalue) is supercritical, while the non-dominant layer remains below its activation threshold and is only weakly driven by interlayer transmission. As a result, activity is effectively confined to the dominant layer, and the two layers display markedly different activity levels.

\item \textbf{Delocalized regimes (AD$_1$ and AD$_2$).}
Delocalized behavior occurs in the regions labeled AD$_1$ and AD$_2$ in Fig.~\ref{fig:phase_diagram_schematic}, where both layers sustain endemic activity and their dynamical behavior becomes comparable. In region AD$_1$, delocalization is induced by increasing the spreading rate $\beta/\mu$ above the activation threshold of the non-dominant layer at weak interlayer coupling. In contrast, region AD$_2$ corresponds to strong interlayer coupling, where increasing $\eta/\beta$ synchronizes the activation of the layers and the system behaves as a single effective network, independently of the individual layer thresholds.
\end{itemize}

The crossover between these two regimes is structural in nature and can be characterized in terms of the spectral properties of the supra-contact matrix. While the transition between the localized and the delocalized state AD$_1$ was characterized in Ref.~\cite{de2017disease}, in Ref.~\cite{tey2024eigenvectorlocalizationuniversalregime}, the transition to the delocalized state AD$_2$ was studied perturbatively and in terms of localization properties of the leading eigenvector. For a two-layer multiplex, this analysis shows that the characteristic scale of the interlayer coupling separating localized and delocalized behavior is set by the difference between the leading eigenvalues of the two layers, $\eta/\beta \sim \Lambda_1 - \Lambda_2$, where $\Lambda_1$ and $\Lambda_2$ are the largest eigenvalues of the individual layers.
 For random regular multiplex networks, exact expressions for the largest eigenvalue and the corresponding eigenvector are available and allow one to locate the crossover more precisely [appendix]. 

In the following sections, we exploit this theoretical framework to analyze node-level autocorrelations and show how they can be used to infer the global state of the multiplex system—namely, whether it is inactive or active, and whether spreading is localized on a single layer or delocalized across layers. We focus in particular on the case of partial observability, and ask whether the state of an unobserved dominant layer can be inferred from the dynamics of a non-dominant one, or vice versa.

\section{Analytical Approximation of the Autocorrelation}
\label{sec:autocorrelation_approx}

In this section, we derive a closed-form analytical approximation for the steady-state autocorrelation of node states in the contact-based social contagion model on multiplex networks. This result constitutes the main analytical contribution of this work and provides the theoretical basis for using autocorrelations as a diagnostic of the global state of a multiplex spreading process. The approximation relies on the mean-field master equation \eqref{eq:mstereq}, which neglects joint infection probabilities and therefore becomes inaccurate in regimes with strong intra- or interlayer correlations. We refer the reader to Ref.~\cite{PhysRevE.97.062309} for closely related derivations in the single-layer SIS case and to Appendix~\ref{sec:corrcalc} for technical details.

The state of a node $u$ in layer $\alpha$ at time $t$ can be described by a two-state Markov process, with $X^\alpha_u(t)=1$ if the node is infected and $X^\alpha_u(t)=0$ otherwise. Under the mean-field approximation, the corresponding transition probabilities follow directly from the master equation \eqref{eq:mstereq} and can be written as
\begin{align}
    X^\alpha_u(t): 0 &\rightarrow 1 \quad \text{with probability } 1 - q^\alpha_u(t), \\
    X^\alpha_u(t): 1 &\rightarrow 0 \quad \text{with probability } \mu - \mu(1 - q^\alpha_u(t)),
    \label{eq:events}
\end{align}
where $q^\alpha_u(t)$ is the probability that node $u$ in layer $\alpha$ does not receive infection from any neighbor at time $t$.

These transitions define the discrete-time transition matrix
\begin{equation}
W =
\begin{bmatrix}
q^\alpha_u & 1 - q^\alpha_u \\
\mu - \mu(1 - q^\alpha_u) & (1 - \mu) + \mu(1 - q^\alpha_u)
\end{bmatrix},
\label{eq:transmat}
\end{equation}
which fully characterizes the effective single-node dynamics within the mean-field approximation.

The steady-state Pearson autocorrelation of node $u$ in layer $\alpha$ at time lag $h$ is defined as
\begin{equation}
\rho_{uu}^{\alpha\alpha}(t,h) =
\frac{\mathbb{E}[X^\alpha_u(t) X^\alpha_u(t+h)] - (p^\alpha_u)^2}
{\mathrm{Var}[X^\alpha_u]},
\label{eq:Evar}
\end{equation}
where $p^\alpha_u=\langle X^\alpha_u\rangle_\infty$ is the steady-state infection probability. Since $X^\alpha_u(t)$ is Bernoulli distributed, $\mathrm{Var}[X^\alpha_u]=p^\alpha_u(1-p^\alpha_u)$. The joint expectation can be expressed in terms of the $h$-step transition probability of the Markov chain (see Appendix~\ref{sec:corrcalc}), yielding
\begin{equation}
\mathbb{E}[X^\alpha_u(t) X^\alpha_u(t+h)]
= \mathbb{P}\!\left(X^\alpha_u(t+h)=1 \mid X^\alpha_u(t)=1\right)\, p^\alpha_u .
\label{eq:prob}
\end{equation}

From this construction one obtains the closed-form expression
\begin{equation}
\rho^{\alpha\alpha}_{uu}(h) =
\frac{
\left[
\frac{(1 - q^\alpha_u) + \mu q^\alpha_u [q^\alpha_u(1 - \mu)]^h}
{1 - q^\alpha_u(1 - \mu)}
\right] p^\alpha_u - (p^\alpha_u)^2
}
{p^\alpha_u - (p^\alpha_u)^2},
\label{eq:corr}
\end{equation}
which gives the steady-state autocorrelation of a node as a function of the recovery probability $\mu$, the steady-state infection probability $p^\alpha_u$, and the probability $q^\alpha_u$ of not receiving infection from neighbors. Equation~\eqref{eq:corr} predicts an exponential decay in the time lag $h$, controlled by the factor $q^\alpha_u(1-\mu)$, and reduces to the known SIS result in the single-layer limit~\cite{PhysRevE.97.062309}.

We now summarize the qualitative behavior implied by Eq.~\eqref{eq:corr} as a function of the position in the phase diagram shown in Fig.~\ref{fig:phase_diagram_schematic}, keeping $\mu$ and the time lag $h$ fixed.

\paragraph*{Inactive phase (I).}
In the inactive region, to the left of the dominant-layer activation threshold $(\beta/\mu)_c^{(1)}$, the only stable solution is $p^\alpha_u=0$ for all nodes and layers. In this regime dynamical fluctuations are absent and autocorrelations are not defined. This region is therefore irrelevant for autocorrelation-based inference and will not be considered further.

\paragraph*{Activation threshold.}
At the activation threshold of the multiplex, $\beta/\mu = 1/\bar{\Lambda}_{\max}$, corresponding to the vertical boundary between regions I and AL/AD in Fig.~\ref{fig:phase_diagram_schematic}, the steady-state infection probabilities become infinitesimally small and $q^\alpha_u \approx 1$. Expanding Eq.~\eqref{eq:corr} to leading order yields, for lag $h=1$,
\begin{equation}
\rho^{\alpha\alpha}_{uu}(1) = 1 - \mu ,
\end{equation}
independently of network structure. At this point infections are rare and short-lived, and recovery dominates the dynamics, resulting in an autocorrelation controlled solely by the recovery probability.

\paragraph*{Active--localized regime (AL).}
In the region labeled AL in Fig.~\ref{fig:phase_diagram_schematic}, the multiplex is endemic because the dominant layer is supercritical, while the non-dominant layer remains below its activation threshold and is only weakly driven by interlayer transmission. In this regime Eq.~\eqref{eq:corr} predicts qualitatively different autocorrelation behavior for the two layers. In the dominant layer, increasing $\beta/\mu$ leads to a growth of $p^\alpha_u$ and a corresponding decrease of $q^\alpha_u$, producing a monotonic decay of the autocorrelation. In contrast, the non-dominant layer exhibits sporadic, externally driven infections and its autocorrelation remains approximately constant at the critical value $1-\mu$. This separation between layer autocorrelations constitutes a robust analytical signature of localization.

\paragraph*{Active--delocalized regime induced by transmissibility (AD$_1$).}
For weak or moderate interlayer coupling, increasing the spreading rate $\beta/\mu$ beyond the activation threshold of the non-dominant layer drives the system into the region labeled AD$_1$ in Fig.~\ref{fig:phase_diagram_schematic}. In this regime both layers sustain endemic activity, and the distinction between dominant and non-dominant layers disappears. Equation~\eqref{eq:corr} predicts that the autocorrelations of both layers decay with increasing $\beta/\mu$, following the same qualitative mechanism observed for the dominant layer in the localized regime. Delocalization in this case is induced by transmissibility rather than by strong coupling.

\paragraph*{Active--delocalized regime induced by coupling (AD$_2$).}
In the region AD$_2$, above the horizontal boundary in Fig.~\ref{fig:phase_diagram_schematic}, strong interlayer coupling synchronizes the dynamics of the layers, causing them to activate simultaneously and behave as a single effective system. In this regime Eq.~\eqref{eq:corr} predicts a concurrent decay of autocorrelations in both layers as $\beta/\mu$ increases. The decay is driven by infection events originating both within layers and across layers, resulting in similar temporal correlation patterns irrespective of the individual layer properties.

\paragraph*{Crossover regime and limitations.}
Near the localization--delocalization boundary separating regions AL and AD$_2$, strong interlayer correlations invalidate the independence assumption underlying the master equation \eqref{eq:mstereq}. As a result, the analytical approximation \eqref{eq:corr} becomes quantitatively inaccurate in this crossover region, particularly in AD$_2$, where interlayer correlations are dominant. We therefore complement the analytical predictions with dynamical simulations and numerical integration of the master equation in Sec.~\ref{sec:results}.

Taken together, these results show that node-level lag 1 autocorrelations provide distinct and analytically predictable signatures of the main regions of the multiplex phase diagram. In particular, a plateau at $1-\mu$ for one layer together with a decaying autocorrelation in the other is indicative of the active--localized regime, while simultaneous decay of autocorrelations in both layers signals delocalized spreading. Importantly, in a partial observability setting, the observation of a layer with relatively low activity and an autocorrelation pinned close to the critical value $1-\mu$ provides a clear signature that the observed layer is non-dominant and is being weakly driven by coupling to another, unobserved layer that is in the supercritical phase. These features form the basis for inferring the global state of the multiplex system—and the activity of an unobserved layer—from partial observations of a single layer, while also delineating the regimes where such inference becomes unreliable.

\section{Numerical and dynamical results}
\label{sec:results}

We now test the analytical predictions of Sec.~\ref{sec:autocorrelation_approx} against dynamical simulations of the underlying stochastic process and numerical integration of the mean-field master equation~\eqref{eq:mstereq}. The goal is twofold: to identify autocorrelation-based signatures of the different regions of the phase diagram (inactive/active and localized/delocalized spreading), and to assess whether the state of an unobserved layer (e.g., a dominant platform with limited access) can be inferred from measurements performed on a single observed layer.

\subsection{Simulation protocol and numerical integration}
We implement discrete-time \emph{synchronous} updating, in which all nodes are updated simultaneously from the configuration at the previous time step. This choice matches the discrete-time master equation~\eqref{eq:mstereq} and enables a direct comparison between simulations and mean-field predictions.

To sample the active regime close to threshold without being trapped in the absorbing configuration ($X^\alpha_u=0$ for all $u$ and $\alpha$), we use the quasi-stationary (QS) method~\cite{deoliveira2004simulatequasistationarystate}. When the process attempts to reach the absorbing state, it is instead reset to a previously stored active configuration. As a consequence, measurements below the true activation threshold should be interpreted with care, since the QS procedure sustains a small residual activity in the subcritical region. In this sense, it provides a phenomenological proxy for sporadic reactivation events that may occur in real-world systems due to external inputs.

Unless stated otherwise, we set $\mu=0.5$. While $\mu=1$ is often used for convenience, under synchronous updating it can induce spurious short-lag anticorrelation effects; we discuss this issue explicitly in Sec.~\ref{sec:mu_problem}.

We focus primarily on a two-layer multiplex composed of random regular (RR) layers with degrees $k_1=30$ and $k_2=10$, which we refer to as dominant and non-dominant layers, respectively. Correlation measurements typically use $N=1000$ nodes per layer, while density curves are computed at $N=5000$ to reduce finite-size fluctuations. Each run lasts $4\times 10^4$ time steps; the first $2\times 10^4$ are discarded as transient, and observables are computed in the stationary window. Numerical predictions are obtained by integrating the master equation~\eqref{eq:mstereq}. Code and scripts are available in Ref.~\cite{Joan2025}.

\subsection{Correlation observables}
Throughout this section we report node autocorrelations $\rho^{\alpha\alpha}_{uu}(h)$, intralayer two-node correlations between nearest neighbors $\rho^{\alpha\alpha}_{uv}(h)$, and interlayer correlations between replicas of the same node $\rho^{12}_{uu}(h)$ (and, for completeness, $\rho^{12}_{uv}(h)$). Unless stated otherwise, we focus on lag $h=1$, for which Sec.~\ref{sec:autocorrelation_approx} provides a simple interpretation in terms of the balance between recovery and infection events.

\begin{figure}
  \centering
  \includegraphics[width=0.9\columnwidth]{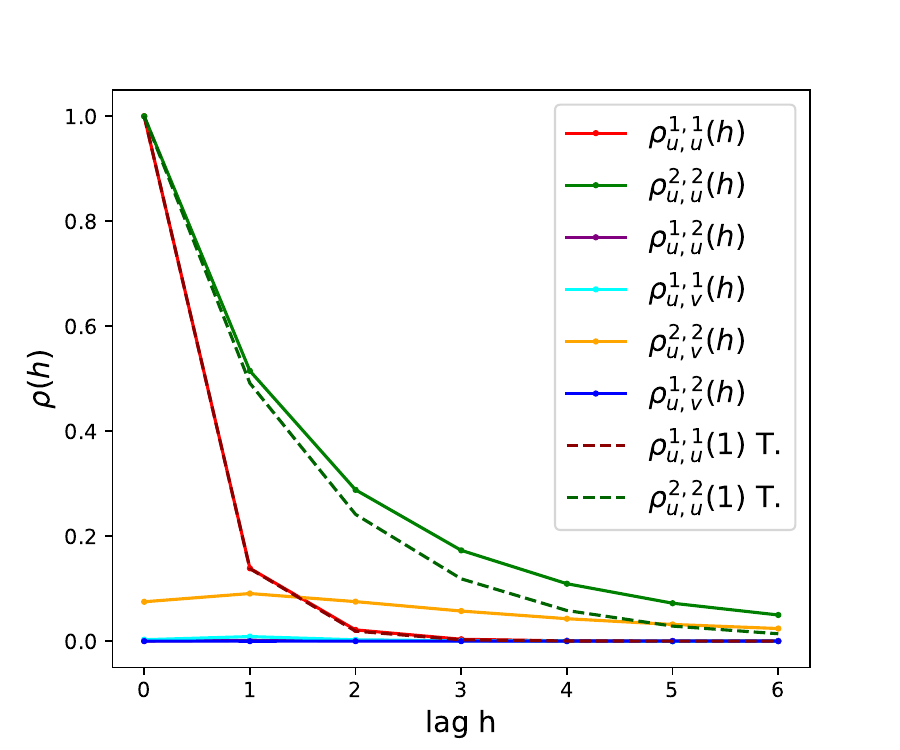}
  \caption{Representative correlation functions at steady state for $\beta/\mu=0.1$, $\eta/\beta=0.01$ and $\mu=0.5$. The quantities $\rho_{u,u}^{1,1}$ and $\rho_{u,u}^{2,2}$ are autocorrelations for layers 1 and 2, while $\rho_{u,v}^{1,1}$ and $\rho_{u,v}^{2,2}$ are nearest-neighbor intralayer correlations. The interlayer correlation between replicas of the same node is $\rho_{u,u}^{1,2}$, and $\rho_{u,v}^{1,2}$ denotes interlayer correlations between different nodes. Curves labelled ``T.'' correspond to mean-field predictions obtained from numerical integration of~\eqref{eq:mstereq}.}
  \label{fig:autocorr}
\end{figure}

Figure~\ref{fig:autocorr} illustrates the typical behavior of the correlation functions once the system has reached stationarity. By definition, $\rho^{\alpha\alpha}_{uu}(h=0)=1$ and the autocorrelation decays with lag $h$. Deviations between dynamical simulations and mean-field predictions are more pronounced in the non-dominant layer, consistent with larger two-node correlations $\rho^{22}_{uv}(h)$ that violate the independence assumption underlying~\eqref{eq:mstereq}. In Appendix~\ref{app:ER} we repeat the same analysis for Erd\H{o}s--R\'enyi layers and obtain the same qualitative picture.

\subsection{Active--localized regime (AL) and crossover to AD$_1$}
\begin{figure*}
    \centering
    \includegraphics[width=1\textwidth]{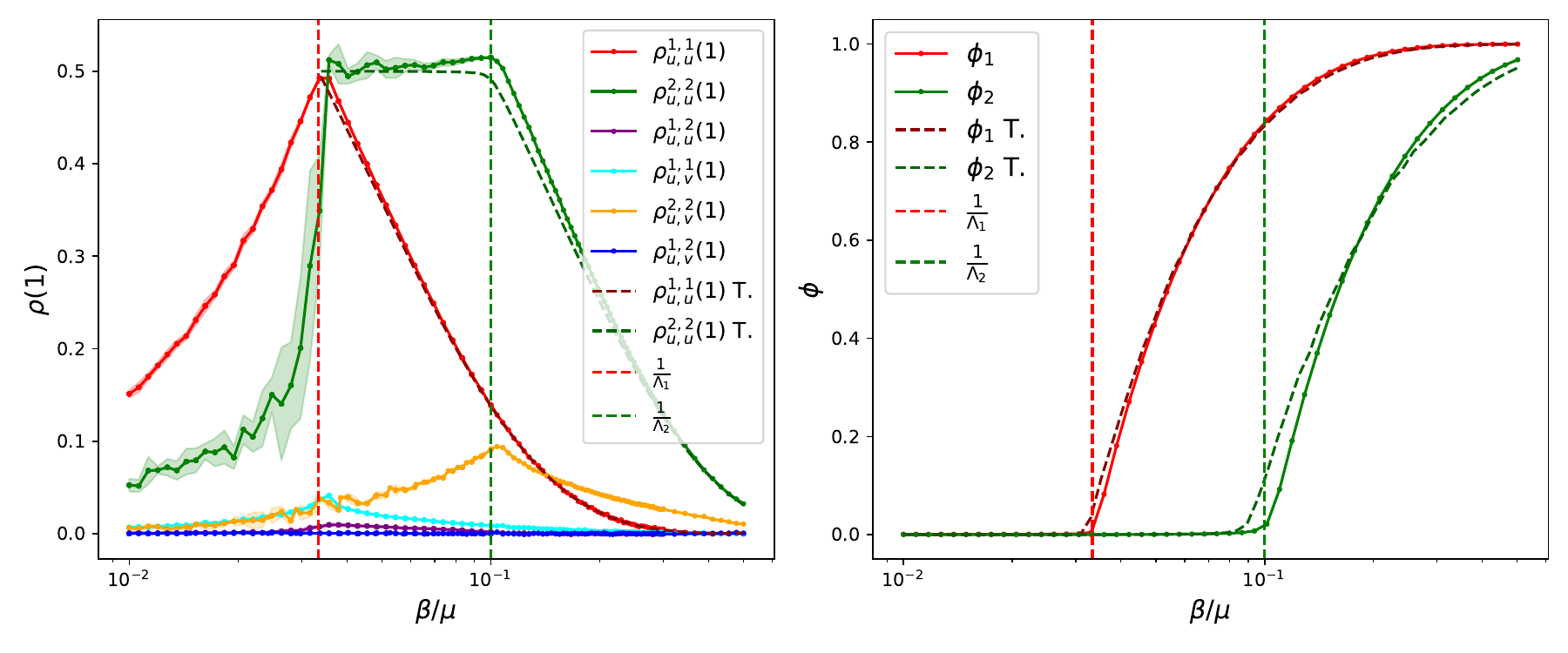}
    \caption{Active--localized regime (AL) for $\eta/\beta=0.01$. Left: correlation measures at lag $h=1$ as functions of $\beta/\mu$. Right: stationary density in each layer as a function of $\beta/\mu$. Vertical dashed lines indicate the single-layer thresholds $1/\Lambda_1$ and $1/\Lambda_2$. Curves labelled ``T.'' correspond to numerical integration of the master equation~\eqref{eq:mstereq}.}
    \label{fig:eta001_N1000}
\end{figure*}

We first consider parameter values in the region AL of the schematic phase diagram in Fig.~\ref{fig:phase_diagram_schematic}, by fixing a weak interlayer coupling $\eta/\beta=0.01$ and varying $\beta/\mu$. Figure~\ref{fig:eta001_N1000} shows lag-$1$ correlations (left) and stationary densities (right). In this regime, two-node correlations between different nodes remain small, and the mean-field master equation provides a good qualitative description of the autocorrelation trends.

Below the true activation threshold of the multiplex, the exact dynamics is absorbed in the inactive state (region I), so autocorrelations are not defined due to the absence of fluctuations. The QS procedure sustains a small residual activity in this region, producing an apparent increase of $\rho^{\alpha\alpha}_{uu}(1)$ as $\beta/\mu$ approaches threshold; we therefore treat this subcritical behavior as a QS artifact and focus on $\beta/\mu$ at and above activation.

As $\beta/\mu$ increases past the multiplex activation point, the dominant layer (layer~1) becomes supercritical while the non-dominant layer remains below its own activation threshold. This is the defining feature of the active--localized regime AL: the multiplex is endemic because layer~1 is active, whereas layer~2 is only weakly driven by interlayer transmission and remains effectively subcritical. Consistent with Sec.~\ref{sec:autocorrelation_approx}, $\rho^{11}_{uu}(1)$ decreases beyond the activation of layer~1, while $\rho^{22}_{uu}(1)$ remains approximately pinned near the critical value $1-\mu$ throughout the AL region. Once $\beta/\mu$ crosses the non-dominant activation threshold, the system leaves AL and enters AD$_1$, and $\rho^{22}_{uu}(1)$ begins to decay as layer~2 sustains endemic activity.

This separation between layer autocorrelations provides a direct diagnostic of AL: one observes a decaying autocorrelation in the dominant layer together with an autocorrelation pinned close to $1-\mu$ and a low activity density in the non-dominant layer. In a partial-observability setting, observing a layer with low activity and $\rho^{\alpha\alpha}_{uu}(1)\simeq 1-\mu$ is therefore consistent with that layer being non-dominant and weakly driven by an unobserved dominant layer that is already supercritical.

Nearest-neighbor intralayer correlations $\rho^{\alpha\alpha}_{uv}(1)$ exhibit peaks close to the activation point of each layer, providing an additional signature of the onset of sustained intralayer spreading. The peak is larger for the non-dominant layer, consistent with its activation occurring at a higher effective spreading rate $\beta/\mu$.

\subsection{Active--delocalized regime induced by coupling (AD$_2$)}
\begin{figure*}
    \centering
    \includegraphics[width=1\textwidth]{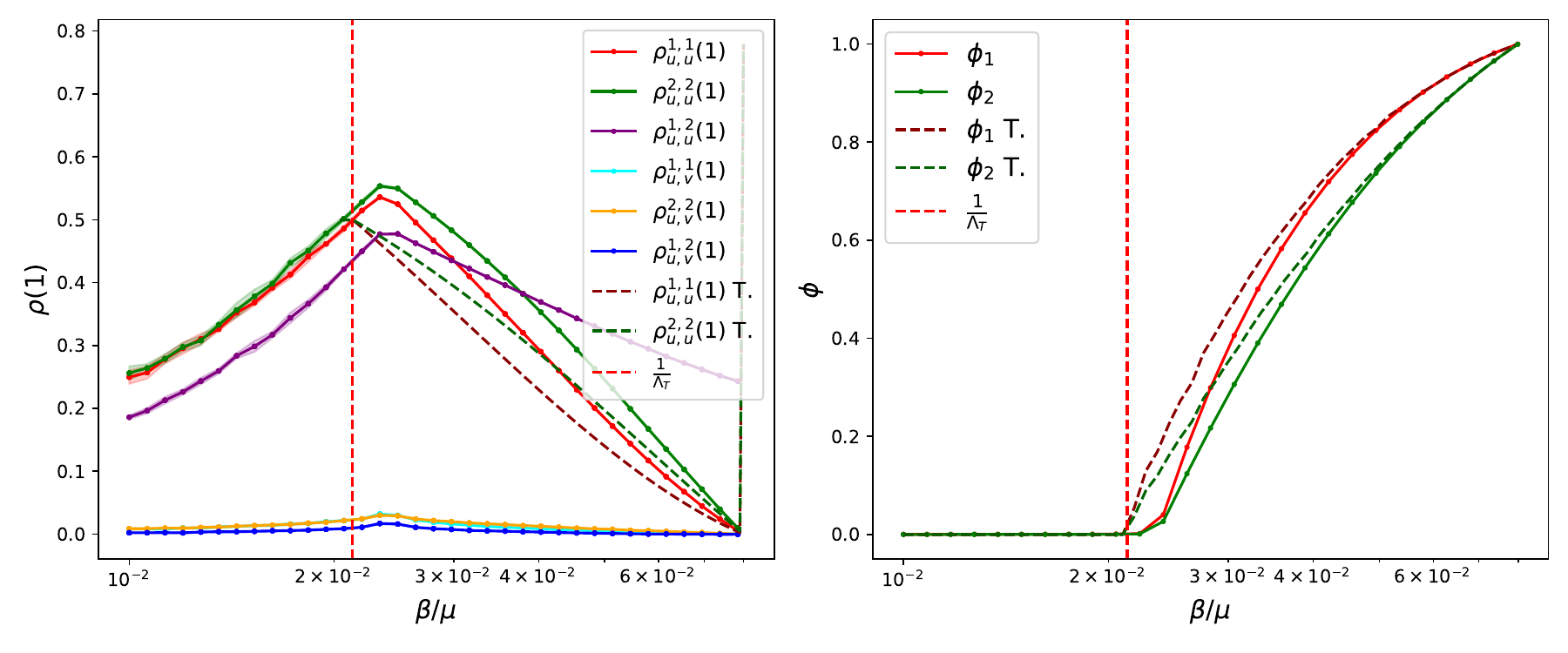}
    \caption{Active--delocalized regime AD$_2$ ($\eta/\beta=25$). Left: correlation measures at lag $h=1$ vs.\ $\beta/\mu$. Right: stationary density in each layer. The vertical dashed line at $1/\Lambda_T$ indicates the spectral activation threshold based on the largest eigenvalue of the supra-contact matrix (Eq.~\eqref{eq:maxeigen}). See caption in the original version for the full list of observables.}
    \label{fig:eta25_N1000}
\end{figure*}

We next consider parameters in the region AD$_2$ of the schematic phase diagram (Fig.~\ref{fig:phase_diagram_schematic}), by setting a strong interlayer coupling $\eta/\beta=25$, well above the localization--delocalization crossover scale. As shown in Fig.~\ref{fig:eta25_N1000}, interlayer correlations between replicas of the same node, $\rho^{12}_{uu}(1)$, become large, while intralayer nearest-neighbor correlations remain comparatively small. This behavior reflects the dominance of interlayer transmission, which strongly synchronizes the states of node replicas across layers.

In this regime, strong interlayer correlations invalidate the independence assumption underlying the mean-field master equation, and the numerical integration of Eq.~\eqref{eq:mstereq} no longer provides a quantitatively accurate description of the autocorrelations. Nevertheless, the activation threshold extracted from simulations is well predicted by the spectral criterion based on the largest eigenvalue of the supra-contact matrix, confirming the validity of the eigenvalue-based threshold estimate.

Beyond activation, both layers become endemic essentially simultaneously, which is the defining feature of the active--delocalized regime AD$_2$. The autocorrelations of both layers decay as $\beta/\mu$ increases, with a faster decay in the layer with larger degree. This reflects the increased number of infection sources in that layer, which enhances temporal randomness in node states. From the perspective of inference under partial observability, regime AD$_2$ is characterized by the absence of a marked separation between layer autocorrelations: both layers exhibit comparable activity levels and jointly decreasing $\rho^{\alpha\alpha}_{uu}(1)$, making it impossible to identify a dominant layer from autocorrelation measurements alone.

\subsection{Active--localized regime near the AL--AD$_2$ boundary (intermediate coupling)}
\begin{figure*}
    \centering
    \includegraphics[width=1\textwidth]{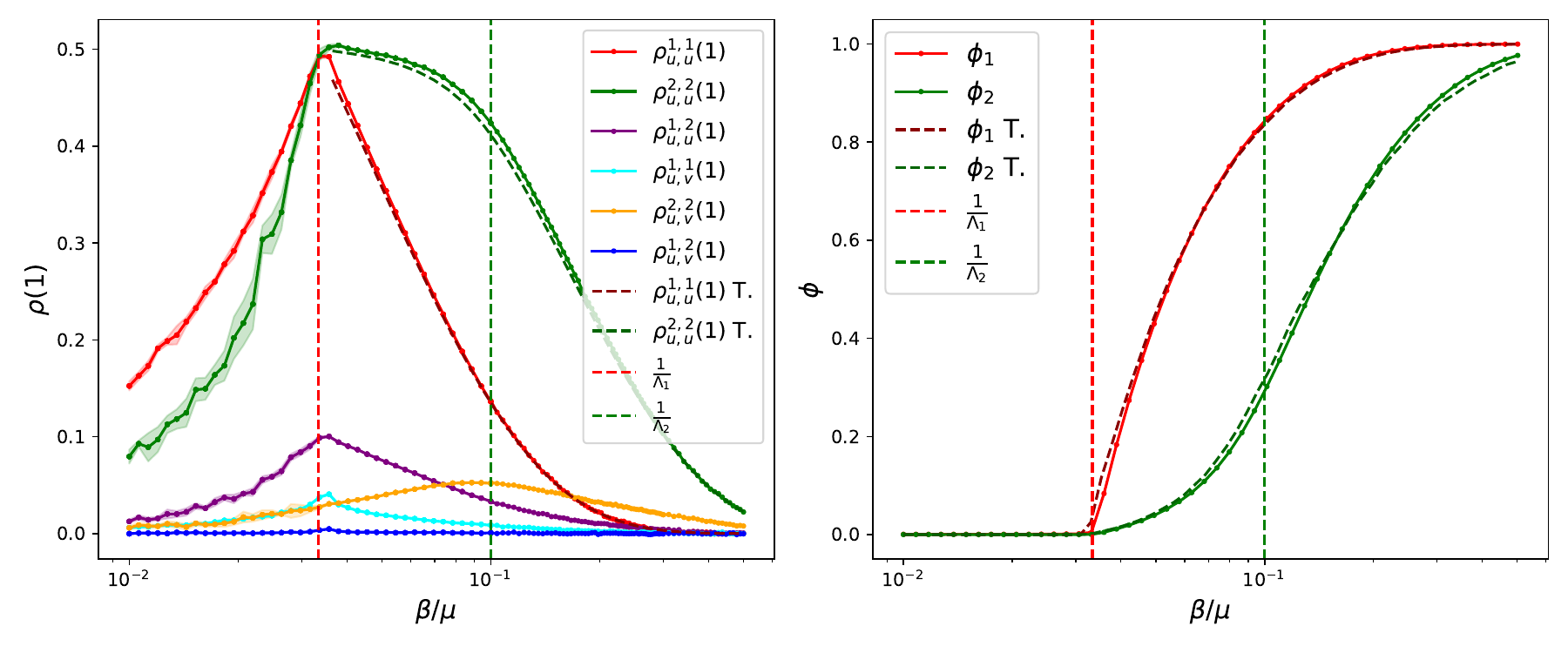}
    \caption{Active--localized regime close to the AL--AD$_2$ crossover ($\eta/\beta=1$). Left: correlation measures at lag $h=1$ vs.\ $\beta/\mu$. Right: stationary density in each layer. Vertical dashed lines indicate the single-layer thresholds $1/\Lambda_1$ and $1/\Lambda_2$. See caption in the original version for the full list of observables.}
    \label{fig:eta1_N1000}
\end{figure*}

We finally consider an intermediate interlayer coupling $\eta/\beta=1$, corresponding to a parameter region still within AL but close to the crossover to AD$_2$ (Fig.~\ref{fig:phase_diagram_schematic}). In this case, the dominant layer is supercritical, while the non-dominant layer remains formally below its single-layer activation threshold. However, sustained seeding from the dominant layer allows the non-dominant layer to maintain a non-negligible density of infected nodes.

As a consequence, $\rho^{22}_{uu}(1)$ may start decreasing before the non-dominant layer reaches its true activation threshold. Although the system is still in the active--localized regime, intralayer infection events in the non-dominant layer already contribute appreciably to the dynamics, increasing temporal randomness and reducing autocorrelation. This behavior reflects proximity to the AL--AD$_2$ boundary, rather than genuine delocalization.

This regime is particularly relevant for inference under partial observability. An observed non-dominant layer may display a sizable activity level while still being largely driven by interlayer transmission, potentially mimicking delocalized behavior if activity alone is considered. In this case, autocorrelation provides complementary information: a decay of $\rho^{22}_{uu}(1)$ that remains weaker than in the dominant layer signals proximity to, but not yet entry into, the delocalized regime AD$_2$.

\subsection{The AL--AD$_2$ crossover}
\begin{figure}
  \centering
  \includegraphics[width=0.9\columnwidth]{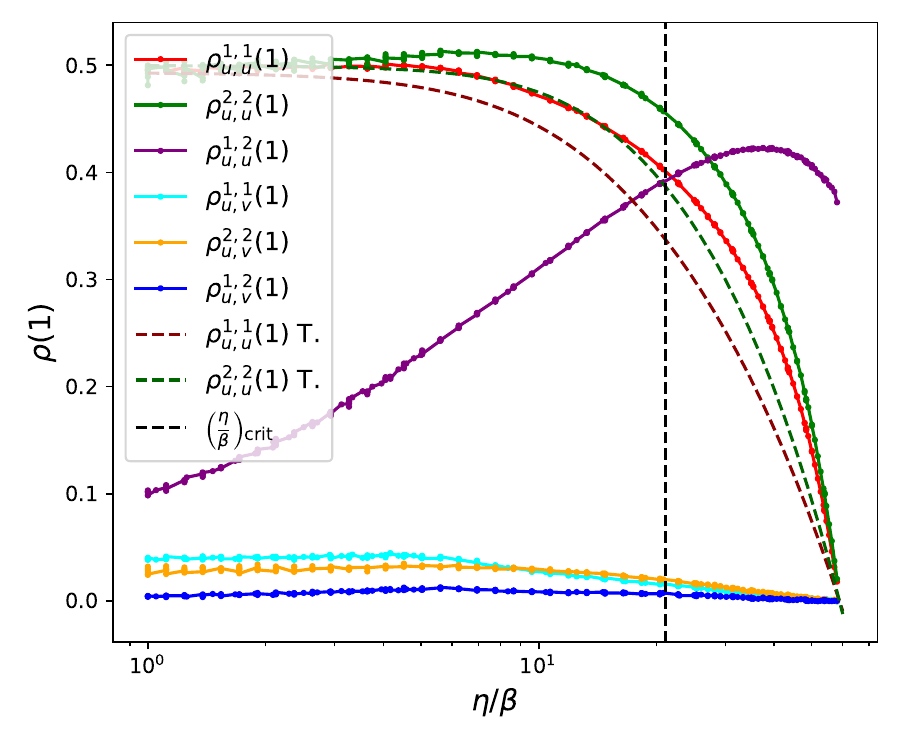}
  \caption{Correlation measures at lag $h=1$ as functions of $\eta/\beta$ at fixed $\beta/\mu=0.034$. The vertical dashed line indicates the AL--AD$_2$ crossover estimated from the structural criterion or the refined IPR-based estimate when available).}
  \label{fig:locdeloc}
\end{figure}

To probe the crossover between the active--localized regime (AL) and the coupling-induced active--delocalized regime (AD$_2$), we fix $\beta/\mu=0.034$ and vary the interlayer coupling $\eta/\beta$. For the RR multiplex with degrees $30$ and $10$, the structural criterion predicts the onset of delocalization at $\eta/\beta \simeq 20$ (Sec.~\ref{sec:model}). Figure~\ref{fig:locdeloc} shows how correlation observables evolve across this scan.

As $\eta/\beta$ increases, the interlayer replica correlation $\rho^{12}_{uu}(1)$ grows, reflecting the increasing synchronization of node states across layers. This correlation reaches a smooth maximum and then decreases as the system enters the AD$_2$ regime, where the layers behave as a single effective system. Importantly, the position of this maximum does not coincide exactly with the structural crossover estimate, and none of the correlation observables exhibit singular behavior. This confirms that the AL--AD$_2$ crossover is structural and smooth, rather than a critical transition, when viewed through temporal correlation diagnostics.

At the same time, discrepancies between dynamical simulations and mean-field predictions increase with $\eta/\beta$, consistent with the progressive breakdown of the node-independence approximation due to strong interlayer correlations in the AD$_2$ regime.

\begin{figure}
  \centering
  \includegraphics[width=0.9\columnwidth]{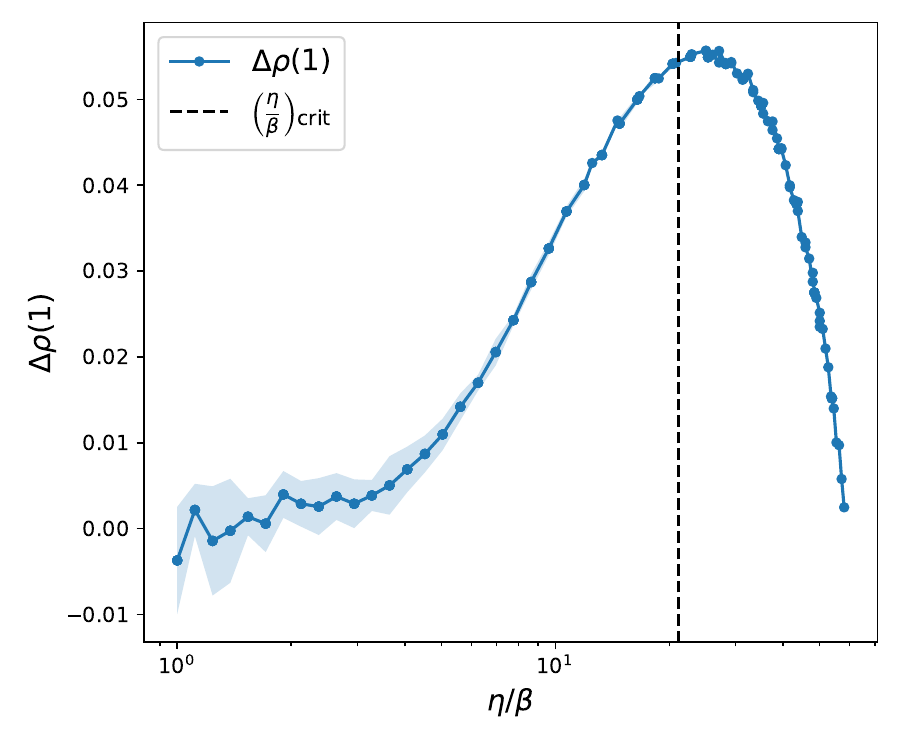}
  \caption{Difference in lag-$1$ autocorrelation between layers, $\Delta\rho(1)=\rho^{22}_{uu}(1)-\rho^{11}_{uu}(1)$, as a function of $\eta/\beta$ at fixed $\beta/\mu=0.034$. The vertical dashed line indicates the AL--AD$_2$ crossover estimated from the structural criterion.}
  \label{fig:diffcorr}
\end{figure}

Motivated by Fig.~\ref{fig:locdeloc}, we also examine the difference
$\Delta\rho(1)=\rho^{22}_{uu}(1)-\rho^{11}_{uu}(1)$ (Fig.~\ref{fig:diffcorr}). This quantity exhibits a smooth maximum in the vicinity of the AL--AD$_2$ crossover, reflecting the gradual reduction of the asymmetry between layer autocorrelations as interlayer coupling increases. While this feature suggests that $\Delta\rho(1)$ may serve as a practical indicator of the crossover in finite systems, we do not interpret it as a universal order parameter. Establishing its robustness across network topologies and parameter regimes would require further investigation.

\subsection{The $\mu$ problem}
\label{sec:mu_problem}

\begin{figure}
  \centering
  \includegraphics[width=0.9\columnwidth]{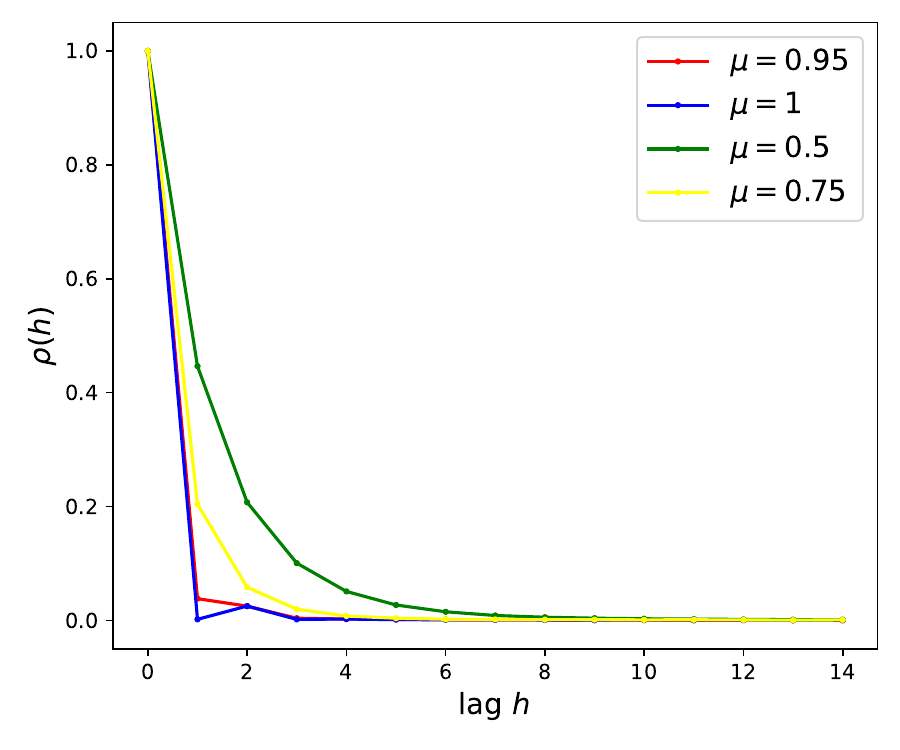}
  \caption{Autocorrelation at different time lags for the dominant layer (degree $30$) coupled to a non-dominant layer (degree $10$), for different values of $\mu$ at fixed $\beta/\mu=0.4$ and $\eta/\beta=1$.}
  \label{eq:muanalysis}
\end{figure}

We conclude by highlighting a technical caveat specific to synchronous updating. At fixed $\beta/\mu=0.4$ and $\eta/\beta=1$, Fig.~\ref{eq:muanalysis} shows that for $\mu$ close to $1$ the measured autocorrelation can exhibit non-monotonic behavior with lag (e.g., $\rho(h=2)>\rho(h=1)$). This is not predicted by the mean-field expression~\eqref{eq:corr}, which yields $\rho^{\alpha\alpha}_{uu}(h>0)=0$ for $\mu=1$, and indicates the presence of higher-order temporal correlations not captured by the node-independence approximation. For this reason, we adopt $\mu=0.5$ as a default choice in the rest of the paper.

\section{Conclusion}

In this work, we addressed the problem of inferring the global state of a multiplex spreading process under conditions of partial observability, a situation that naturally arises in contemporary digital ecosystems where access to platform-level data is uneven. Focusing on a contact-based contagion model on multiplex networks, we showed that node-level temporal autocorrelations provide informative and analytically tractable signatures of the main dynamical regimes of the system.

By combining a mean-field analytical approximation with dynamical simulations, we demonstrated that autocorrelations capture the distinction between inactive, active--localized (AL), and active--delocalized (AD$_1$ and AD$_2$) regimes in a way that is consistent with the underlying phase diagram. In particular, the separation between layer autocorrelations in the AL regime and their simultaneous decay in delocalized regimes provide robust indicators of whether activity is confined to a dominant layer or spread across layers. Crucially, these signatures remain accessible even when only a single layer is observed, enabling inference about the state of unobserved components of the system.

At the same time, our results delineate the limitations of correlation-based inference. Near the localization--delocalization crossover and in strongly coupled regimes, interlayer correlations invalidate the independence assumptions underlying mean-field descriptions, reducing quantitative accuracy and blurring regime boundaries. These limitations highlight the inherently structural nature of the crossover and the need for caution when interpreting temporal correlations in highly coupled multiplex systems.

More broadly, our findings suggest that temporal correlations constitute a powerful yet lightweight diagnostic tool for multiplex dynamics, complementing structural analyses and offering practical advantages in data-limited settings. We expect that extending this approach to empirical data, heterogeneous layer structures, and more general dynamical processes will further clarify the role of partial observability in complex, multi-platform systems.

\begin{acknowledgments}
We acknowledge financial support from the Spanish grant PID2021-128005NB-C22, and from Generalitat de Catalunya under project 2021-SGR-00856.
\end{acknowledgments}

\section*{Appendix}

\subsection{Autocorrelation Calculation}
\label{sec:corrcalc}

In this appendix, the autocorrelation function $\rho_{uu}^{\alpha\alpha}(h)$ is derived following a procedure analogous to that used in Appendix A of \cite{PhysRevE.97.062309}. 

Given the transition matrix $W$, Eq.~\eqref{eq:transmat}, the probability vector describing the state of node $u$ at layer $\alpha$ and time $t$ evolves as
\begin{equation}
    P_u^\alpha(t+h) = P^\alpha_u(t)\cdot W^h,
\end{equation}
where $P^\alpha_u(t) = 
    \begin{bmatrix} 
        1 - p^\alpha_u(t), & p^\alpha_u(t) 
    \end{bmatrix}$,
and $p^\alpha_u(t)$ is the probability that node $u$ at layer $\alpha$ is active at time $t$.

To compute $W^h$, it is convenient to diagonalize $W$. If $W = A W_D A^{-1}$, the evolution becomes
\begin{equation}
    P^\alpha_u(t+h) = P^\alpha_u(t) \cdot A W_D^h A^{-1}.
    \label{eq:PtransA}
\end{equation}

The eigenvalues of $W$ are:
\begin{equation}
    \lambda_1 = 1, \quad \lambda_2 = q(1 - \mu),
\end{equation}
with corresponding eigenvectors:
\begin{equation}
    v_1 = 
    \begin{pmatrix}
        \mu q \\ 1 - q
    \end{pmatrix}, 
    \quad
    v_2 = 
    \begin{pmatrix}
        1 \\ -1
    \end{pmatrix}.
\end{equation}

From these, let the coordinates transformation matrix and its inverse be
\begin{equation}
    \begin{matrix}
        A = \begin{pmatrix}
        \mu q&
        1-q \\
        1 &
        -1
    \end{pmatrix} & A^{-1}= \begin{pmatrix}
        \frac{1}{1-q(1-\mu)} & 
        \frac{1-q} {1-q(1-\mu)} \\ 
        \frac{1}{1-q(1-\mu)}&
        \frac{-\mu q}{1-q(1-\mu)}
    \end{pmatrix}
    \end{matrix}
\end{equation}

Thus, raising $W$ to the $h$-th power yields
\begin{equation}
    W^h = 
    \begin{pmatrix}
        \frac{\mu q + (1 - q)[q(1 - \mu)]^h}{1 - q(1 - \mu)} & 
        \frac{(1 - q)[1 - [q(1 - \mu)]^h]}{1 - q(1 - \mu)} \\[6pt]
        \frac{\mu q[1 - [q(1 - \mu)]^h]}{1 - q(1 - \mu)} & 
        \frac{(1 - q) + \mu q [q(1 - \mu)]^h}{1 - q(1 - \mu)}
    \end{pmatrix}.
    \label{eq:FiWh}
\end{equation}

Each entry of $W^h$ corresponds to a conditional probability:
\begin{equation}
    W^h =
    \begin{bmatrix}
        P_{00} & P_{10} \\
        P_{01} & P_{11}
    \end{bmatrix},
\end{equation}
where $P_{ab} = P\left( X^\alpha_u(t+h) = a \mid X^\alpha_u(t) = b \right)$, with $a, b \in \{0, 1\}$. Here, $X^\alpha_u(t)$ denotes the state of node $u$ at layer $\alpha$ and time $t$, with $X^\alpha_u(t) = 1$ if infected and $X^\alpha_u(t) = 0$ otherwise.

For a binary state variable $X^\alpha_u(t) \in \{0,1\}$, the joint expectation reads
\begin{equation}
    \mathbb{E}[X^\alpha_u(t) X^\alpha_u(t+h)] = P_{11} \cdot p_u^\alpha,
\end{equation}
where $P_{11} = P(X^\alpha_u(t+h) = 1 \mid X^\alpha_u(t) = 1)$ is the conditional probability of node $u$ at layer $\alpha$ being active at time $t+h$ given that it was active at time $t$, and $p^\alpha_u = P(X^\alpha_u(t) = 1)$ is the stationary activation probability.

Using this, the autocorrelation function at lag $h$ is computed as
\begin{equation}
    \rho_{uu}^{\alpha\alpha}(h) = \frac{ \mathbb{E}[X^\alpha_u(t) X^\alpha_u(t+h)] - (p^\alpha_u)^2 }{ p^\alpha_u - (p^\alpha_u)^2 },
\end{equation}
where $p^\alpha_u = P(X^\alpha_u(t) = 1)$ is the stationary activation probability. Substituting the explicit form of $\mathbb{E}[X^\alpha_u(t) X^\alpha_u(t+h)]$ gives
\begin{equation}
    \rho_{uu}^{\alpha\alpha}(h) = 
    \frac{
        \left[ \frac{(1 - q^\alpha_u) + \mu q^\alpha_u [q^\alpha_u(1 - \mu)]^h}{1 - q^\alpha_u(1 - \mu)} \right] p^\alpha_u - (p^\alpha_u)^2
    }{
        p^\alpha_u - (p^\alpha_u)^2
    }.
    \label{eq:corrapp}
\end{equation}

\subsection{Maximum eigenvalue and eigenvector derivation}
\label{app:eigen}

In this appendix, we show the derivation of the exact largest eigenvalue $\bar{\Lambda}_{max}$, and its corresponding eigenvector for a random regular multiplex network. The eigenvector of the largest eigenvalue has a two block structure. By applying the supra-contact matrix, Eq.~\eqref{eq:Rbar}, to this eigenvector, the following system of equations is obtained:
\begin{align}
    \Lambda_1 \cdot v_1 + \epsilon \cdot v_2 &= \bar{\Lambda}_{\max} \cdot v_1, \label{eq:sp1} \\
    \Lambda_2 \cdot v_2 + \epsilon \cdot v_1 &= \bar{\Lambda}_{\max} \cdot v_2, \label{eq:sp2} \\
    N \cdot v_1^2 + N \cdot v_2^2 &= 1, \label{eq:norm}
\end{align}
where $\epsilon = \frac{\eta}{\beta}$ and $N$ is the number of nodes per layer. Eq.~\eqref{eq:norm} characterizes the normalization property of the eigenvector.

For simplicity, we introduce the rescaled variables $\bar{v}_1 = \sqrt{N} \cdot v_1$ and $\bar{v}_2 = \sqrt{N} \cdot v_2$, transforming the normalization condition into $\bar{v}_1^2 + \bar{v}_2^2 = 1$.
By applying this change in Eq. \eqref{eq:norm} and then substituting it in Eq. \eqref{eq:sp1} and Eq.~\eqref{eq:sp2} one finds the solution for $\bar{v}_i$  as
\begin{equation}
    \bar{v}_i^2 = \frac{\epsilon^2}{\epsilon^2 + \left(\Lambda_{\max} - \Lambda_i\right)^2},
    \label{eq:vi}
\end{equation}
for $i \in \{1, 2\}$.

Substituting Eq.~\eqref{eq:vi} into the normalization condition yields
\begin{equation}
    \frac{\epsilon^2}{\epsilon^2 + \left(\bar{\Lambda}_{\max} - \Lambda_1\right)^2} + \frac{\epsilon^2}{\epsilon^2 + \left(\bar{\Lambda}_{\max} - \Lambda_2\right)^2} = 1.
\end{equation}

Rearranging and simplifying this expression leads to a quadratic equation for $\bar{\Lambda}_{\max}$:
\begin{equation}
    \epsilon^2 = \Lambda_{\max}^2 - \bar{\Lambda}_{\max} \left(\Lambda_1 + \bar{\Lambda}_2\right) + \Lambda_1 \Lambda_2.
\end{equation}
\vspace{30pt}

The solution for $\bar{\Lambda}_{\max}$, valid under the assumption that the first layer is dominant ($\Lambda_1 \geq \Lambda_2$), is
\begin{equation}
    \bar{\Lambda}_{\max} = \frac{\Lambda_1 + \Lambda_2 + \sqrt{ \left( \Lambda_1 - \Lambda_2 \right)^2 + 4 \epsilon^2 }}{2}.
    \label{eq:maxeigen}
\end{equation}
\vspace{1pt}

\subsection{Erdö-Renyi simulations}
\label{app:ER}
\begin{figure}
    \centering
    \includegraphics[width=0.9\columnwidth]{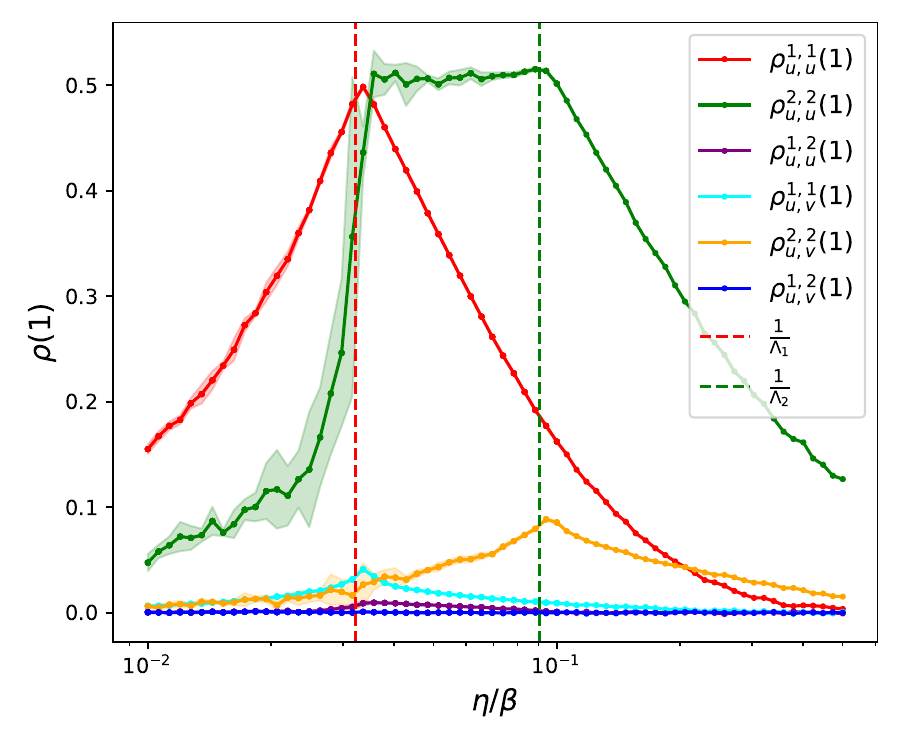} 
    \caption{Correlation measures as functions of $\frac{\beta}{\mu}$ with a set $\frac{\eta}{\beta}=0.01$. The quantities $\rho_{u,u}^{1,1}$ and $\rho_{u,u}^{2,2}$ represent the autocorrelations for layers 1 and 2, respectively, including error bars. The terms $\rho_{u,v}^{1,1}$ and $\rho_{u,v}^{2,2}$ correspond to the correlations between the first neighbors within each layer. The vertical dashed lines at $\frac{1}{\Lambda_1}$ and $\frac{1}{\Lambda_2}$ are the inverse of the largest eigenvalue of each layer's adjacency matrix. The interlayer correlation between the same node is denoted $\rho_{u,u}^{1,2}$, while $\rho_{u,v}^{1,2}$ represents the interlayer correlation between different nodes. Results are averaged over simulations on an Erdö-Renyi (ER) multiplex network with $N = 1000$ nodes, where layer 1 has degree $30$ and layer 2 has degree $10$.}
    \label{fig:EReta001_N1000}
\end{figure}

In this appendix, we present the same set of figures shown in Section~\ref{sec:results}, but for Erdős–Rényi networks instead of random regular networks.

\begin{figure}[]
    \centering
    \includegraphics[width=0.9\columnwidth]{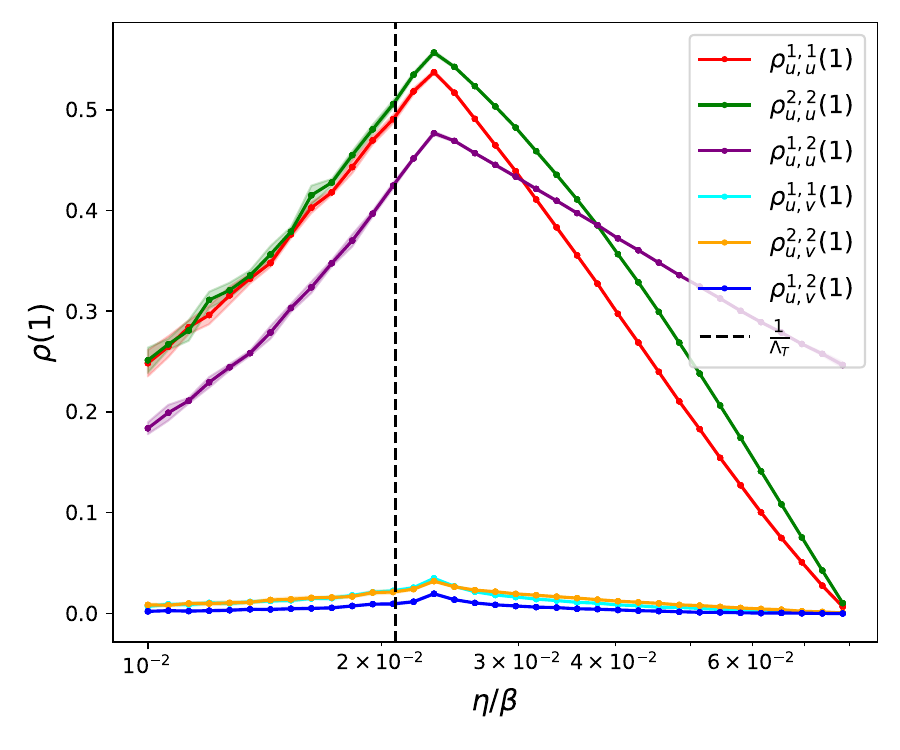} 
    \caption{Correlation measures as functions of $\frac{\beta}{\mu}$ with a set $\frac{\eta}{\beta}=25$. The quantities $\rho_{u,u}^{1,1}$ and $\rho_{u,u}^{2,2}$ represent the autocorrelations for layers 1 and 2, respectively, including error bars. The terms $\rho_{u,v}^{1,1}$ and $\rho_{u,v}^{2,2}$ correspond to the correlations between first neighbors within each layer. The vertical dashed line at $\frac{1}{\Lambda_T}$ is the inverse of the theoretical largest eigenvalue, Eq.\eqref{eq:maxeigen}, considering $\frac{\eta}{\beta}=25$.The vertical dashed line at $\frac{1}{\Lambda_S}\approx43$ is the moment at which the transition occurs in simulations. The interlayer correlation between the same node is denoted $\rho_{u,u}^{1,2}$, while $\rho_{u,v}^{1,2}$ represents the interlayer correlation between different nodes. Results are averaged over simulations on a Erdö-Renyi (ER) multiplex network with $N = 1000$ nodes, where layer 1 has degree $30$ and layer 2 has degree $10$.}
    \label{fig:EReta25_N1000}
\end{figure}

\begin{figure}
    \includegraphics[width=0.9\columnwidth]{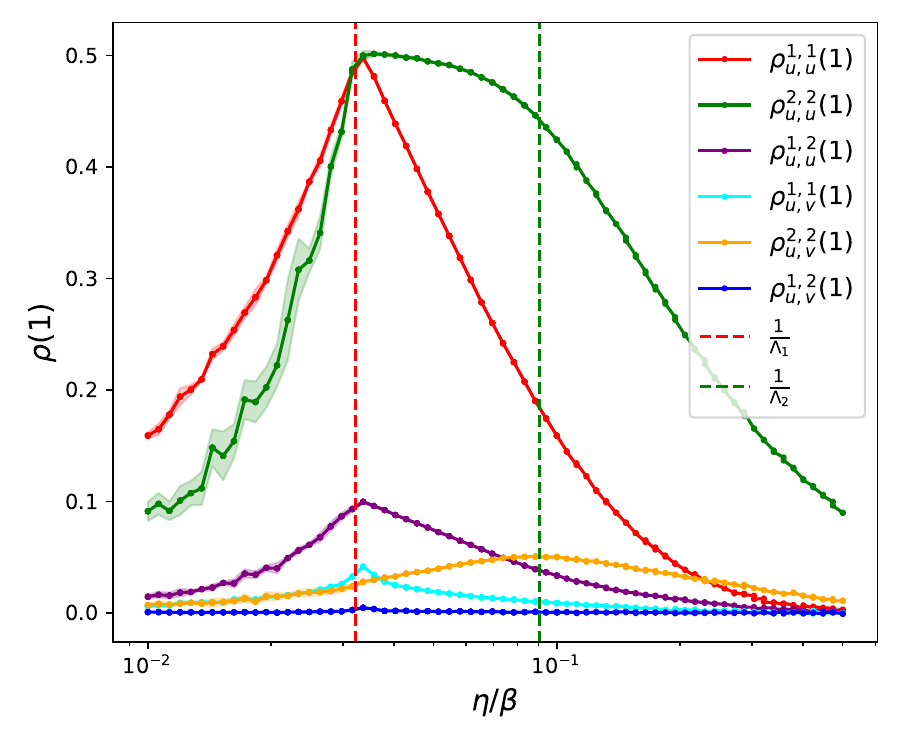} 
    \caption{Correlation measures as functions of $\frac{\beta}{\mu}$ with a set $\frac{\eta}{\beta}=1$. The quantities $\rho_{u,u}^{1,1}$ and $\rho_{u,u}^{2,2}$ represent the autocorrelations for layers 1 and 2, respectively, including error bars. The terms $\rho_{u,v}^{1,1}$ and $\rho_{u,v}^{2,2}$ correspond to the correlations between first neighbors within each layer. The vertical dashed lines at $\frac{1}{\Lambda_1}$ and $\frac{1}{\Lambda_2}$ are the inverse of the largest eigenvalue of each layer's adjacency matrix. The interlayer correlation between the same node is denoted $\rho_{u,u}^{1,2}$, while $\rho_{u,v}^{1,2}$ represents the interlayer correlation between different nodes. Results are averaged over simulations on a Erdö-Renyi (ER) multiplex network with $N = 1000$ nodes, where layer 1 has degree $30$ and layer 2 has degree $10$.}
    \label{fig:EReta1_N1000}
\end{figure}

From Fig.~\ref{fig:EReta001_N1000}, Fig.~\ref{fig:EReta1_N1000}, and Fig.~\ref{fig:EReta25_N1000}, it can be concluded that the correlation behavior observed for Erdős–Rényi networks follows the same qualitative pattern as the random regular network case analyzed in Section~\ref{sec:results}, although the rate at which the autocorrelation decays and the height of the peak may vary slightly.

\bibliographystyle{apsrev4-2}
\bibliography{ref}

\end{document}